\documentclass[prd,onecolumn,nofootinbib,aps,tightenlines,preprintnumbers]{revtex4-1}
\pdfoutput=1

\def\be{\begin{equation}}
\def\ee{\end{equation}}
\def\bea{\begin{eqnarray}}
\def\eea{\end{eqnarray}}

\newcommand{\del}{\partial}

\usepackage{framed}
\usepackage[compat=1.0.0]{tikz-feynman}
\usepackage{multirow}
\usepackage{braket}

\usepackage{color}

\usepackage{amsfonts}
\usepackage{amssymb}
\usepackage{amsthm}
\usepackage{slashed}
\usepackage{bm}
\usepackage[makeroom]{cancel}
\usepackage{amsmath, mathtools}
\usepackage{framed}

\usepackage{graphicx}

\usepackage[margin=0.8in]{geometry}
\usepackage[margin=1cm,font=footnotesize]{caption}

\newcommand{\Lagr}{\mathcal{L}}
\newcommand{\Ord}{\mathcal{O}}

\newcommand{\pref}[1]{(\ref{#1})}
\newcommand{\barr}[1]{\overline{#1}}
\newcommand{\eq}[1]{\begin{align} #1 \end{align}}

\newcommand{\units}[1]{\ensuremath{\mathrm{#1}}}

\newcommand{\mev}{\units{MeV}}
\newcommand{\gev}{\units{GeV}}

\newcommand{\cm}{\units{cm}}

\newcommand{\s}{\units{s}}
\newcommand{\sr}{\units{sr}}

\newcommand{\yr}{\text{yr}}

\begin{document}

\preprint{\hbox{PREPRINT UCI-TR-2021-19}  }

\title{
Light Dark Matter Through the Quark Axial-Vector Current Portal
}
\author{Dillon Berger}
\affiliation{Department of Physics and Astronomy, University of
California, Irvine,
  CA 92697, USA}

\begin{abstract}
We consider a model in which dark matter is a light ($350~\mev$) Dirac fermion which couples to quarks as an axial-vector current.
At center-of-mass (CM) energies below the confinement scale, such a DM interaction structure may be described through its low energy effective couplings to mesons. In this paper we focus on dark matter annihilation into these final state mesons, and its resulting photon spectrum. We present a coarse-grained approach for analyzing our model in which we argue the meson spectrum solely from kinematics and symmetries and thereby estimate the total DM annihilation cross section that corresponds to a given expected photon signal. We then corroborate and refine these findings with the more systematic approach of the chiral Lagrangian description of the model, since the CM energies we consider admit such a perturbative treatment. In both cases, we obtain constraints on the model by comparing the estimated photon signals to current and future observations, and show
that the axial-vector DM portal is significantly more conducive to photon production for lighter forms of dark matter than a pure vector-like portal; establishing itself as a prime candidate for indirect detection probes with significant discovery reach.
\end{abstract}
\maketitle

\section{Introduction}

Recently, there has been significant interest in models of dark matter (DM) in
which the dark matter particle
has a mass $m_\chi \lesssim {\cal O}(\gev)$.
This is appealing because such models evade nearly all constraints on DM imposed by direct detection
experiments, since direct detection experiments are insensitive to the small recoil energies characteristic of $m_\chi\lesssim \Ord(\gev)$. If this sub-GeV dark matter annihilates or decays into SM particles, then the energies of these final-state particles are produced with energies in the $\Ord(\mev)$ range. In the case that these final state particles decay into photons, such a process would in principle be astrophysically observable. In this case, one could then detect the indirect signatures of dark matter by probing dark matter dominated regions of space and searching for an excess of these photons. In this paper, we will derive this photon signal directly from the model and determine bounds imposed on the model's parameters by current diffuse background emission and future observations of dwarf spheroidal galaxies (dSphs).\footnote{ We note that the current experimental sensitivity to detect  $\Ord(\mev)$-photons is relatively poor, but there are currently a number of proposed experiments (see \cite{DeAngelis:2017gra,Caputo:2017sjw,APT}) in development which will improve sensitivity in these regions.}

Dwarf spheroidal galaxies are currently believed to be among the most dark matter dominated regions in the universe \cite{dmDominated}, with mass-to-light ratios of the order $\Ord(10^2 M_\odot)$ \cite{massToLight, massToLight2, massToLight3, massToLight4}. As such, dSphs provide a unique and fertile testing field for searches of indirect signals predicted by theories of dark matter which couple to the standard model. In this paper we will derive constraints imposed on our dark matter model by future indirect DM searches of dSphs. Specifically, we will consider these constraints as they apply to a future indirect search of the Draco galaxy.

For the reasons above, there has been recent interest in sub-GeV dark matter which specifically couples to quarks ~\cite{Boddy:2015efa,Boddy:2015fsa,Boddy:2016fds,Bartels:2017dpb,Cata:2017jar,
Dutra:2018gmv,Kumar:2018heq}. This is because at low energies, all of the ingredients for indirect detection are present: the DM will directly couple to neutral pions (and other mesons; instead of free quarks) through contact interactions, and these mesons will then decay to photons leaving behind a characteristic photon spectrum.

 In this paper, we will consider MeV-range cold dark matter which couples to quarks as an axial-vector current. Previous work has been done on this \cite{Kumar:2018heq} at CM energies $\sim \Ord(\gev)$, but considers only the dominant $\pi^0 \pi^0 \eta$ process (which is inaccessible at the CM energies we consider). Moreover, the analysis therein is carried out at CM energies close to the confinement scale ($\Lambda_{QCD}$),  and so does not lend itself to a perturbative analysis and  provides loose upper bounds on the DM branching ratios and not the model parameters themselves. As noted before, at CM energies near $\Lambda_{QCD}$ there are many additional kinematically accessible final mesonic final states which decay to photons one has to consider. As a result, the net photon signal becomes dominated by the spectrum produced by the $\pi^0 \pi^0 \eta$ process, and so while the bounds obtained on the effective cross sections found in \cite{Kumar:2018heq} are of coarse still true, they become much more stringent for the dark matter masses we consider coupling to quarks as an axial-vector.

In this work we take dark matter to be cold and have mass $m_\chi \sim 350~\mev$, and so the CM energy of the annihilation process is $\sqrt{s}\sim 2m_\chi \sim 700~\mev$. Further, we make no assumptions about the flavor structure of the DM couplings to the quarks. One of the more salient properties of dark matter coupling as an axial-vector to dark matter is how it differs from other portals, such as the vector-like portal \cite{ourPaper}. As we will see, the meson spectrum characteristic of the axial couplings differs drastically from that of the vector-current coupling, in such a way that it is significantly more conducive to photon production at lower DM masses---making it a prime candidate for indirect detection probes.

The plan of this paper is as follows.  In Section II, we present an  overview of the model we consider, introducing the coupling of dark matter the quark axial-vector current. In Section \ref{spectrum} we argue the meson spectrum that results from our model by considerations of quantum numbers. In Section \ref{branchingRatios} we outline a useful general method of estimating the resultant photon spectrum, using only the knowledge we have of the final states (i.e. without mention to a Lagrangian description). Next, in Section \ref{chiralTheory} we outline and justify the chiral Lagrangian description of the model, the technical details of which can be found in Appendix \ref{chiralAppendix}. Then in Section \ref{observations} we go over our methods for estimating the photon background and expected experimental signal strength for DM annihilation photon production for current and future observations. Finally, in Section \ref{results} we report our findings on the photon signal predicted by our theory and the constraints thereby imposed on the DM branching ratios and the model's parameter space.

\section{The Model}

We consider a model in which dark matter annihilates through a coupling to an axial-vector quark current.
In this model,
we take dark matter to be a Dirac fermion ($\chi$). The axial-vector coupling to the quarks is given by 
\bea
{\cal L}_{int} &=&  \sum_q {\alpha_q \over \Lambda^2}(\bar{\chi}\gamma^\mu \gamma^5 \chi)
( \bar{q}\gamma_\mu \gamma^5 q ).
\label{eq:IntLagrangian2}
\eea

\noindent It will be useful to denote the axial-vector current which couples to the quarks $q$ with strength coupling $\alpha_q$ as 

\bea
a^\mu = {1 \over \Lambda^2}(\bar{\chi}\gamma^\mu \gamma^5 \chi),
\eea

\noindent where $\Lambda$ is the expected mass of a new massive particle that would mediate the DM-quark interaction.
Note that, in general, this interaction does not respect the $SU(3)_L \times SU(3)_R$ chiral symmetry of massless QCD. However if the $\alpha_q$ couplings do not discriminate between the quarks then this interaction does preserve the approximate low energy $SU(3)_F$ flavor symmetry. As mentioned before, we take the mass of the dark matter to be light at $m_\chi\sim 350 $ MeV, which yields a CM energy $\sqrt{s}\sim 700~\mev$ well below the confinement scale. Hence we do not expect dark matter to annihilate into freely propagating quarks, but rather into kinematically accessible mesonic final states.
\section{Spectrum}
\label{spectrum}

 We now consider the possible mesonic final states of this model. Now, not only do the final states need to be kinematically accessible, they also need to have quantum numbers consistent with the initial state of the dark matter's wavefunction. That is, the final states must have quantum numbers consistent with a spin-1 axial-vector. We can therefore determine the allowed SM final states by imposing the fact that they must have $J^{PC} = 1^{++}$ \cite{Kumar:2013iva}. Now, recall that both the $P$-parity and $C$-parity of a particle are factored into an intrinsic $C/P$-parity, and an extrinsic $C/P$-parity which depends on the angular quantum numbers of the wavefunction. Further recall that the overall $C/P$-parity of a \textit{state} is given by the product of of the intrinsic $C/P$ parities of each particle composing the state times the overall phase contribution from the extrinsic $C/P$-parity. For a boson, the extrinsic $P$-parity $(\mathcal{P})$ and extrinsic $C$-parity $(\mathcal{C})$ of a boson is given by\footnote{ We use $\mathcal{C}, \mathcal{P}$ to denote extrinsic parities of a state and $C, P$ to denote the overall parities of a state.}  

\eq{
\mathcal{C}: (-1)^{L+S} \qquad \qquad \mathcal{P}:(-1)^L.  \label{parity}
}

So, kinematic constraints notwithstanding, the dark matter can decay into any $J=1$ quark bound state with appropriate $P$ and $C$. This immediately implies that there can be no single pseudoscalar meson final states, and also no single vector meson final states, since the vector mesons all have $J^P = 1^-$. So if the vector mesons are to appear in final states, they either need to come in pairs or with at least one pseudoscalar meson. However, at the CM energies we consider ($\sqrt{s}\sim 700~\mev$) such vector states are kinematically forbidden, and only become relevant at $\sqrt{s} \gtrsim m_\pi + m_\rho \sim 914~\mev$. It is insightful to compare the final-state spectrum of the axial-vector coupling to dark matter to that of a pure vector coupling, which we now do.

\begin{table}[h!]
\begin{tabular}{|c|c|c|}
\hline
\text{Final State $(1^{--})$} & $E_{tot}$ \ \text{(MeV)} \\
\hline
 $\pi^+ \pi^-$ & 280  \\
 $\rho^0$ & 775\\
 $\omega$ & 782\\
   $\rho^\pm \pi^\mp $ & 915 \\
$K^+ K^-$ & 986 \\
 $K_L K_S$ & 994 \\
   $\phi$ & 1020 \\
\hline
\end{tabular}
\ \ \ \ 
\begin{tabular}{|c|c|c|}
\hline
\text{Final State $(1^{++})$} & $E_{tot}$ \ \text{(MeV)} & \text{Order} \\
\hline
$\pi^0 \pi^0 \pi^0$ & 405 & $p^3$\\
 $\pi^0 \pi^+ \pi^-$ & 412 & $p+ p^3$ \\
   $(\pi^0)^3 \pi ^- \pi ^+$ & 680 & $p + p^3$ \\
 $\pi^0 (\pi ^-)^2 (\pi ^+)^2$ & 690 & $p+p^3$\\
\hline
\end{tabular}
\caption{({\bf Right}) All kinematically accessible mesonic final states with quantum numbers consistent with the DM axial-current, and the order in the chiral Lagrangian at which they occur}
\label{table}
\end{table}

In order for the final-state quantum numbers to be consistent with the dark matter's initial state, the vector states must be in the $|L,S\rangle = |1,0\rangle$ state and that the axial vector states in the $|L,S\rangle = |1,1\rangle$ state. Each of the vector and axial-vector final states share the same overall phase $(-1)$ due to extrinsic parity. However since the intrinsic parity of pseudoscalar and vector mesons is odd, we must have an odd number of of final state mesons for the axial case, and an even number for the vector case. Considering the kinematic constraints at $\sqrt{s}\sim700~\mev$, we see that dark matter coupling to the quarks as an axial-vector can only admit final states containing an odd number of pseudoscalar mesons. The vector case is very similar except that the kinematic constraints are far more restrictive insofar as photon production is concerned, and that one is forced into an even number of mesons (as opposed to odd).

In Table \ref{table} (left) we have listed the five lightest final states that result from the dark matter coupling to quarks as a vector current, and on the right we have done the same for the axial-vector current. Recalling that $\pi^0\to 2\gamma$ almost always, and that $\pi^+\pi^-$ nearly never produces photons (i.e. $\lesssim 10^{-5}\ $\%), it is immediately clear that the axial-vector coupling is far more conducive to photon production for lighter forms of dark matter, since in this case the lightest photon-producing final state happens at $\sqrt{s}\sim 405~\mev$ (producing six photons), as opposed to the vector case in which the lightest state is $915~\mev$ which produces only two photons. Specifically, for the DM mass we consider the vector-current coupling admits zero photon-producing final states and the axial-current coupling admits four of them---all of these states are shown in Table \ref{table} (right).

Lastly, it is important to note that, not only are there four photon-producing final states through the axial-vector portal (as opposed to the zero via the vector portal). What is more is that Table \ref{table} (right) is an exhaustive list of \textit{all} of the accessible final states through the axial-vector portal at $\sqrt{s} \sim 700~\mev$. This is because the only accessible final states are combinations of $3,5,7\ldots$ mesons whose net mass is less than $700~\mev$. Scanning over all combinations of mesons the reader may verify that those listed in Table \ref{table} (right) are the only such final states. 

Lastly, we note a convenient consequence of the set of all mesonic final states for the axial-vector case. Namely, since there are no vertices for which the dark matter current couples to an even number of mesons, this means that the tree-level Feynman rules for the DM-meson contact interactions are protected from all 1-loop corrections with mesons running the loop.

In summary, we assume the dark matter is cold, and so we may take the dark matter to be approximately at rest. We take the dark matter to have mass $m_\chi = 350~\mev$ so that the CM energy of the DM annihilation process is $\sqrt{s} \sim 2 m_\chi\sim 700~\mev$. At this energy only combinations of an odd number ($>1$) of the scalar mesons whose mass is less than $\sqrt{s}\sim 700~\mev$ are kinematically accessible. In particular, we find that the only final states with the correct quantum numbers which are also kinematically accessible are the $\pi^0 \pi^+ \pi^-$, $\pi^0 \pi^0 \pi^0$, $\pi^0 \pi^0 \pi^0\pi^+ \pi^-$ and $\pi^0 \pi^+ \pi^- \pi^+ \pi^-$ states, and that the tree-level Feynman rules for these processes are protected from 1-loop corrections with mesons running the loop.
\

\section{Dark Matter Branching Ratios}
\label{branchingRatios}

To obtain coarse estimates of the total DM annihilation cross section in a Lagrangian-free way, we consider constraints placed on the the dark matter branching ratios into the allowed mesonic final states. We do this by incorporating the observations of diffuse photon emission as well as observations from possible future measurements of the Draco dSph galaxy. In this section we will treat the branching ratios as free parameters and make no mention of the underlying particulars of the model.

Let us first define the effective average cross section $\langle \sigma v \rangle_{eff}$, as the sum over the cross sections through each channel $\langle \sigma v \rangle_c$,  weighted by the average number of photons produced by the final state $N_c$, given that process $c$ occurs. In terms of the branching ratios, the effective cross section is then given by 

\begin{align}
\langle \sigma v \rangle _{eff} &= \sum_{c \ {\rm channels}} \langle \sigma v \rangle_c N_c \\
&= \langle \sigma v \rangle_{tot.} \sum_{c \ {\rm channels}} {\rm BR}(\barr{X}X \to c) N_c
\end{align}

\noindent where $\langle \sigma v\rangle_{tot.}$ is the total dark matter annihilation cross section.

The total photon signal $N_S$ is given by 

\eq{
N_S &= \frac{1}{8\pi m_\chi^2} \bar{J}^{ann.} (I_{exp}\Delta \Omega)  \sum_{c \ {\rm channels}}\langle \sigma v \rangle_c N_c \\
&= \frac{\langle \sigma v \rangle_{tot.}}{8\pi m_\chi^2} \bar{J}^{ann.} (I_{exp}\Delta \Omega)  \sum_{c \ {\rm channels}} {\rm BR}(\barr{X}X \to c) N_c 
}

\noindent where $\bar{J}^{ann}$ is the average DM annihilation $J$-factor of the target, $I_{exp}$ is the exposure and $\Delta \Omega$ is the solid angle subtended by the target. One can easily compute $N_c$ in the parent particle frame for each channel, and then boosting the resulting spectrum back to the lab frame. The details by which we do this are outlines in Appendix \ref{boostdecayspectrumappendix}, and the results of these calculations are shown in Table \ref{table2}. Now, to calculate the photon signal $N_S$ exactly is a model dependent endeavor, since one would need to calculate the DM branching ratios into each mesonic final state $c$. However, we can treat the branching ratios and total annihilation cross section as free parameters and  obtain an estimate on its bound, which is what we will do in Section \ref{results}. It will be convenient to define 

\eq{
\Theta_{diff} &:= \bar{J}^{ann}_{diff} \frac{I_{exp}\Delta\Omega_{diff}}{8\pi m_\chi^2}\sim 1.4 \times 10^{33} ~ {\rm cm}^{-3} ~ {\rm {s}} \\
\Theta_{draco} &:= \bar{J}^{ann}_{draco} \frac{I_{exp}\Delta\Omega_{draco}}{8\pi m_\chi^2}\sim 2.2 \times 10^{29} ~ {\rm cm}^{-3} ~ {\rm {s}}
}

\noindent and label the states by an index (shown in Table \ref{table2}). Note that since $\sum_c {\rm BR}_c \leq 1$  it follows that $\sum_c N_c {\rm BR}_c  < \sum_c N_c:= N_c^{tot.}$, and so the maximum photon signal $N_S^{max}$ is a pure function of the total cross section

\eq{
N_S^{max} = \Theta  N_c^{tot.} \langle \sigma v\rangle_{tot}   \label{NsMax}
}

\noindent In Section \ref{results} we use \pref{NsMax} to obtain a general bound on the total DM annihilation cross section for both Draco and the diffuse emission.

%

\section{The Application of Chiral Perturbation Theory to Dark Matter Interactions With
Axial Currents}
\label{chiralTheory}

 Since
the coupling of dark matter to the light mesons is fundamentally governed by QCD and the dark
matter-quark current contact
interaction, we expect this interaction to exist at lower energies as well wherein the DM couples to the quark-confined states (mesons). At the CM energies we consider it is viable to express this interaction
using the chiral Lagrangian formalism,
in which the dark matter appears as an axial-vector spurion $a_\mu$ which, in general, breaks the approximate Standard Model flavor symmetry. We follow a similar approach to that followed in \cite{Kumar:2018heq, ourPaper}.

\begin{table}[h!]
\begin{tabular}{|c|c|c|}
\hline
\text{Channel $c$ } & $N_c$  & index\\
\hline
 $\pi^0 \pi^+ \pi^-$ & 0.53 & 1\\
$\pi^0 \pi^0 \pi^0$ & 1.58 & 2\\
 $\pi^0 (\pi ^-)^2 (\pi ^+)^2$ & 0.83 & 3\\
  $ (\pi^0)^3 \pi ^- \pi ^+$ & 2.42 & 4\\
\hline
\end{tabular}
\caption{Table of all accessible dark matter annihilation final states $c$. The $N_c$ column is the average number of observed photons, given that the final state on the left is produced. The index is simply to assign an arbitrary ordering to the final states $c$.}
\label{table2}
\end{table}

Now, in order to describe the dark matter axial interaction with the quarks at energies below the confinement scale, we need to write down all operators allowed by the symmetries. We use the chiral Lagrangian formalism, which is just a convenient parametrization of all such operators. Our expansion parameter is $p/\Lambda_{QCD} \sim 0.7$, which isn't as small as one might dream. Therefore, we will also consider orders of the chiral Lagrangian beyond lowest order (\ref{lowestOrder}). 

At lowest order, the chiral Lagrangian is

\begin{equation}
\mathcal{L}^{(2)}=\frac{f^{2}}{4} \operatorname{Tr}\left[D_{\mu} U\left(D^{\mu} U\right)^{\dagger}\right]\label{lowestOrder}
\end{equation}

\noindent where

\begin{align}
D_{\mu} U &=\partial_{\mu}U - i\{a_\mu, U \} \\
U &=e^{\frac{i \sqrt{2}}{f} \Phi} \\
\Phi=\frac{\lambda^{a}}{\sqrt{2}} \pi^{a}(x) &=\left(\begin{array}{ccc}
\frac{\pi^{0}}{\sqrt{2}}+\frac{\eta_{8}}{\sqrt{6}} & \pi^{+} & K^{+} \\
\pi^{-} & -\frac{\pi^{0}}{\sqrt{2}}+\frac{\eta_{8}}{\sqrt{6}} & K^{0} \\
K^{-} & \bar{K}^{0} & -\frac{2 \eta_{8}}{\sqrt{6}}
\end{array}\right)\\
a^\mu &\equiv
\left(
                \begin{array}{ccc}
                  a^\mu_u & 0 & 0 \\
                  0 & a^\mu_d & 0 \\
                  0& 0 & a^\mu_s\\
                \end{array}
              \right) .
\label{eq:PhiLag}
\end{align}

\noindent and the pion decay constant is $f \sim 92~\mev$. Expanding $U$ in equation (\ref{lowestOrder}) we obtain the order $p$ and $p^2$ contributions to the Feynman rules for the final states listed in Table \ref{table}. The Feynman rules thereby obtained at this order are given below.

The next highest order chiral Lagrangian contains the order $p^3$ and $p^4$ contributions to the Feynman rules, and is given by

\begin{align}
\mathcal{L}^{(4)} &=L_{1} \operatorname{tr}\left[\left(D_{\mu} U\right)^{\dagger} D_{\mu} U\right]^{2} \nonumber \\
&+L_{2} \operatorname{tr}\left[\left(D_{\mu} U\right)^{\dagger} D_{\nu} U\right] \operatorname{tr}\left[\left(D^{\mu} U\right)^{\dagger} D^{\nu} U\right] \nonumber \\
&+L_{3} \operatorname{tr}\left[\left(D_{\mu} U\right)^{\dagger} D^{\mu} U\left(D_{\nu} U\right)^{\dagger} D^{\nu} U\right]\nonumber \\
& -i L_{9} \operatorname{tr}\left[V^{\mu \nu}\left[\left(D_{\mu} U\right)^{\dagger} D_{\nu} U -D_{\mu} U\left(D_{\nu} U\right)^{\dagger}\right]\right] \label{secondOrder}
\end{align}

\noindent where 

\begin{align}
V_{\mu\nu} = \del_\mu a_\nu - \del_\nu a_\mu
\end{align}

\noindent and $L_1,L_2, L_3, L_9$ are phenomenologically measured dimensionless constants of order $\Ord(10^{-3})$. 

Now, this Lagrangian $\Lagr^{(4)}$ only corrects processes involving 3 or more final-state mesons, since at lowest order in $\Phi$ we have that $\mathcal{O}(|DU|^2) \sim \mathcal{O}(\Phi^2) + \mathcal{O}(a_\mu \Phi)$ and hence  $\mathcal{O}(|DU|^4) \sim \mathcal{O}(\Phi^4) + \mathcal{O}(a_\mu \Phi^3)$, where we have dropped terms of order $\mathcal{O}(a_\mu^2)$. We note also that, at lowest order we get $\mathcal{O}(|DU|^6) \sim \mathcal{O}(\Phi^6) + \mathcal{O}(a_\mu \Phi^5)$, so the next order chiral Lagrangian will indeed contain corrections to our 5-meson final states. These corrections goes as $ C_6 \left( p /\Lambda_{QCD} \right)^5 \sim  C_6 \cdot  0.15$, where $C_6$ is a constant of dimension $[-2]$ introduced by the $\mathcal{O}(p^6)$ chiral Lagrangian $\Lagr^{(6)}$. It has been shown in \cite{constants} that these constants are of order $\mathcal{O}(C_6) \sim 10^{-8}\mev^{-2}$, which therefore gives us license to ignore these $\mathcal{O}(p^6)$ corrections. For the remainder of this paper, we take Lagrangains \pref{lowestOrder} and \pref{secondOrder} to be the complete description of a dark matter axial current coupled to quarks at $\sqrt{s} \sim 700$ MeV. We note that, contrary to \cite{ourPaper} it's not necessary to treat the DM branching fractions as free parameters, although we do this analysis anyhow. Rather, in our case  the dark matter is sufficiently light to justify a perturbative treatment of the chiral Lagrangian and take the tree level vertices as reliable estimates of the Feynman rules.

The details of the expansion of the chiral Lagrangian into its constituent meson fields and their corresponding Feynman rules are given in Appendix \ref{chiralAppendix}. Some notable results from this expansion are that each of our Lagrangians is directly proportional to $\alpha_{ud}/\Lambda^2$, and hence the Feynman rule corresponding to each channel is also proportional to $\alpha_{ud}/\Lambda^2$. Obtaining a coupling of this form is to be more-or-less expected from the symmetry and kinematic arguments we made in Section \ref{spectrum}. Kinematically, the coupling $\alpha_s$ corresponds to the dark matter interactions with the heavier scalar mesons (i.e. those composed of at least one strange quark) which are inaccessible at the CM energies we consider and so we expect all of the relevant couplings to go as $\alpha_u \pm \alpha_d$.

\section{Photon Spectra \& Comparison To Observations}
\label{observations}

In certain models or at higher CM energies like those in \cite{ourPaper},\cite{Kumar:2018heq}, the photon spectrum can be the result of a multistep decay processes of intermediate mesons (e.g. Kaons, Vector Mesons) before finally producing photons. 
In our case however, the only kinematically accessible final states are those listed in Table \ref{table}, which are all DM contact interactions. Consequently, the only accessible photon producing final state meson is the $\pi^0$, which decays to two photons essentially 100\% of the time. We neglect the $\pi^\pm$ photon production since the $\pi^\pm \to \gamma$ branching ratio essentially zero ($\lesssim 10^{-5}$~\%).

\begin{figure}[t]
\centering
\includegraphics[scale=0.65]{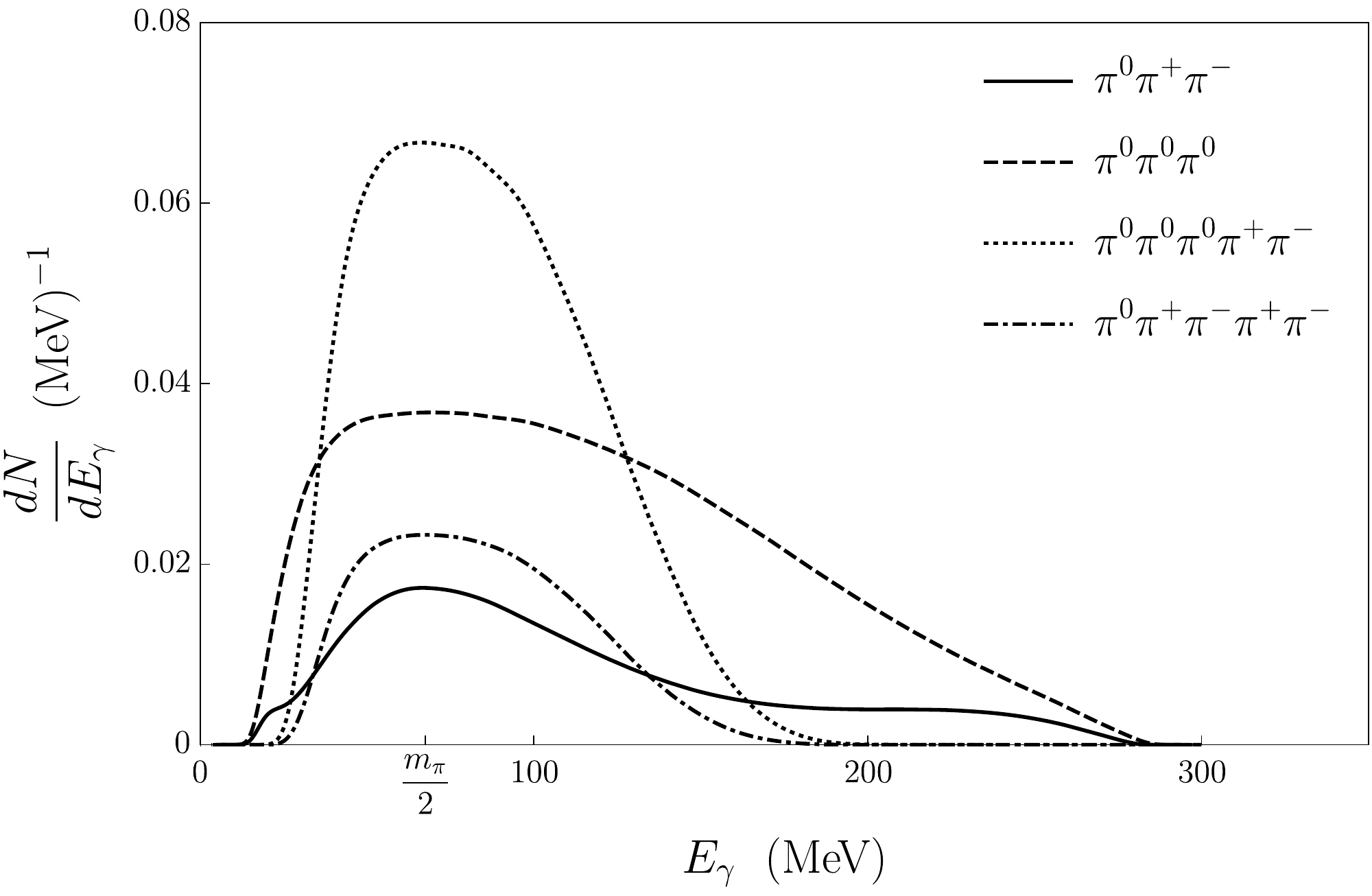}
\caption{Photon spectrum for all kinematically accessible final states. The center of mass energy has been taken to be 700 MeV.
\label{photonSpectrum}
}
\end{figure}

In Figure~\ref{photonSpectrum} we show the
photon spectrum
obtained through the production of neutral pions produced from dark
matter annihilation, assuming that
$\sqrt{s} \sim 700$ MeV. The individual spectra have been normalized to the number of photons produced (i.e. two times the number of pions). In our calculations each spectrum is
weighted according to their branching fractions, though this is not shown in Figure~\ref{photonSpectrum}. As is discussed in Appendix
\ref{boostdecayspectrumappendix}, the photons produced by $\pi^0$ decay
yield a signal with a peak at $m_{\pi} / 2$ which smear out when boosting back to the lab frame, which can be seen in Figure \ref{photonSpectrum}.

 Following~\cite{Kumar:2018heq}, we consider
constraints on the model
 from observations of diffuse photon emission, and from future observations of
photon emission from the Draco dSph.
We will take as a benchmark, an experiment with a fractional $1 \sigma$ energy resolution of
$\epsilon = 0.3$
	and an exposure of $I_{exp}=3000~\cm^2~\yr$.

For the diffuse emission, we restrict our attention
to latitudes greater than $ 20^\circ$. In this region, and in the energy range
$0.8~\mev - 1~\gev$, the isotropic flux observed by COMPTEL and EGRET can be well
fit~\cite{Boddy:2015efa,Strong:2004de} to
the function
\bea
\frac{d^2 \Phi^{iso.}}{d\Omega~dE_{obs.}} &=& 2.74 \times 10^{-3}
\left( \frac{E_{obs.}}{\mev} \right)^{-2.0} \cm^{-2} \s^{-1} \sr^{-1} \mev^{-1}.
\label{eq:ObservedDiffuseFlux}
\eea

\noindent The expected number of observed events $N_O$ between the energies $E_-$ and $E_+$ is therefore given by
\bea
N_{O} &=& 8.6 \times 10^{4} \left(\frac{\mev}{E_-} - \frac{\mev}{E_+} \right)
\frac{(I_{exp.} \Delta \Omega)}{\cm^2~\yr~\sr} .
\eea

\noindent Since our predicted signal peak is at $m_{\pi}/2$ , we will consider a bin of photon energies centered at $E_0 = m_\pi/2$. and a width of $0.3 E_0$ That is, we consider an energy bin between $E_-= (m_{\pi}/2)(1-\epsilon)$
and $E_+=(m_{\pi}/2)(1+\epsilon)$.
 
 To get an estimate on the constraints imposed on our model from diffuse emission, we impose that within this energy bin the number of expected signal events $N_S$ not exceed the number of observed events. That is, we impose that we must have that $N_S < N_O$ in order for the model to be consistent with current observations from diffuse emission.

Now, when obtaining constraints that would be derived from future observations of a dense dark matter region such as the Draco dSph, the relevant quantity to consider is a statistically significant observation of the model's predicted value of $N_S$. As noted in Section \ref{observations}, even in the absence of a photon signal due to DM annihilation there are background photon fluctuations of order $\sigma \sim \sqrt{N_O^{draco}}$, and so it makes sense take this to be our estimated uncertainty. We can therefore rule out a model that predicts a signal $N_S$ at $5\sigma$ confidence if in a future observation no such $N_S > 5\sigma$ excess is detected. Absent of new information about the background emission that could improve resolution, this automatically renders searches for models which predict $N_S < 5\sigma$ inconclusive. If however such a $N_S > 5\sigma$ photon excess is detected upon future observations of Draco, then using \pref{NmaxDraco} we can obtain an estimate on the minimum total DM annihilation cross section consistent with such an observation. Together these constraints provide upper and lower bounds and hence a range of cross sections consistent with a $5\sigma$ Draco photon excess and the diffuse emission.


\subsection{Calculation of DM annihilation photon signal $N_S$}

In this section we briefly outline how we go about calculating the predicted photon signal from DM annihilation $N_S$ from the total DM annihilation cross section, for both the diffuse emission and the Draco case. In general differential photon flux from dark matter annihilation is given by 

\begin{align}
\frac{d^2 \Phi}{d\Omega~dE_\gamma} = \frac{\langle \sigma v\rangle }{8 \pi m_\chi^2} \bar{J}^{ann} \frac{dN_\gamma}{dE_\gamma}
\end{align}

\noindent where $\bar{J}$ is the average $J$-factor of the target for dark matter annihilation, which accounts for the geometry and DM distribution of the target. The average $J$-factor for diffuse emission \cite{Cirelli:2010xx} and for Draco \cite{GeringerSameth:2011iw} are given by 

\begin{align}
\bar{J}_{d i f .}^{a n n}&=3.5 \times 10^{21} \ \mathrm{GeV}^{2} \mathrm{~cm}^{-5} \mathrm{sr}^{-1}\\
\bar{J}_{Draco }^{a n n .}&=6.94 \times 10^{21} \  \mathrm{GeV}^{2} \mathrm{~cm}^{-5} \mathrm{sr}^{-1}
\end{align}

We account for instrumental energy resolution by convolving the injected photon spectrum
with a Gaussian smearing function 
\bea
R_\epsilon (E_{obs.},E_\gamma) &=& \frac{1}{\sqrt{2\pi} \epsilon E_\gamma}
\exp \left(-\frac{(E_{obs.}-E_\gamma)^2}{2\epsilon^2 E_\gamma^2} \right) .
\eea

\noindent Now, in terms of the exposure $I_{exp}$ and the solid angle of the target $\Delta \Omega$, the number of photons expected within the energy window $E_- \leq E_{obs} \leq E_+$ is therefore given by 

\begin{align}
N_S = \frac{\Xi}{8\pi m_\chi} \bar{J}^{ann} (I_{exp} \Delta \Omega) \int_{E_-}^{E_+} dE_{obs} \int_0^\infty dE_\gamma \frac{dN_\gamma}{dE_\gamma} R_\epsilon(E_{obs}, E_\gamma)
\end{align}

\section{Results}
\label{results}


\begin{figure}[t!]
\centering
\includegraphics[scale=.5]{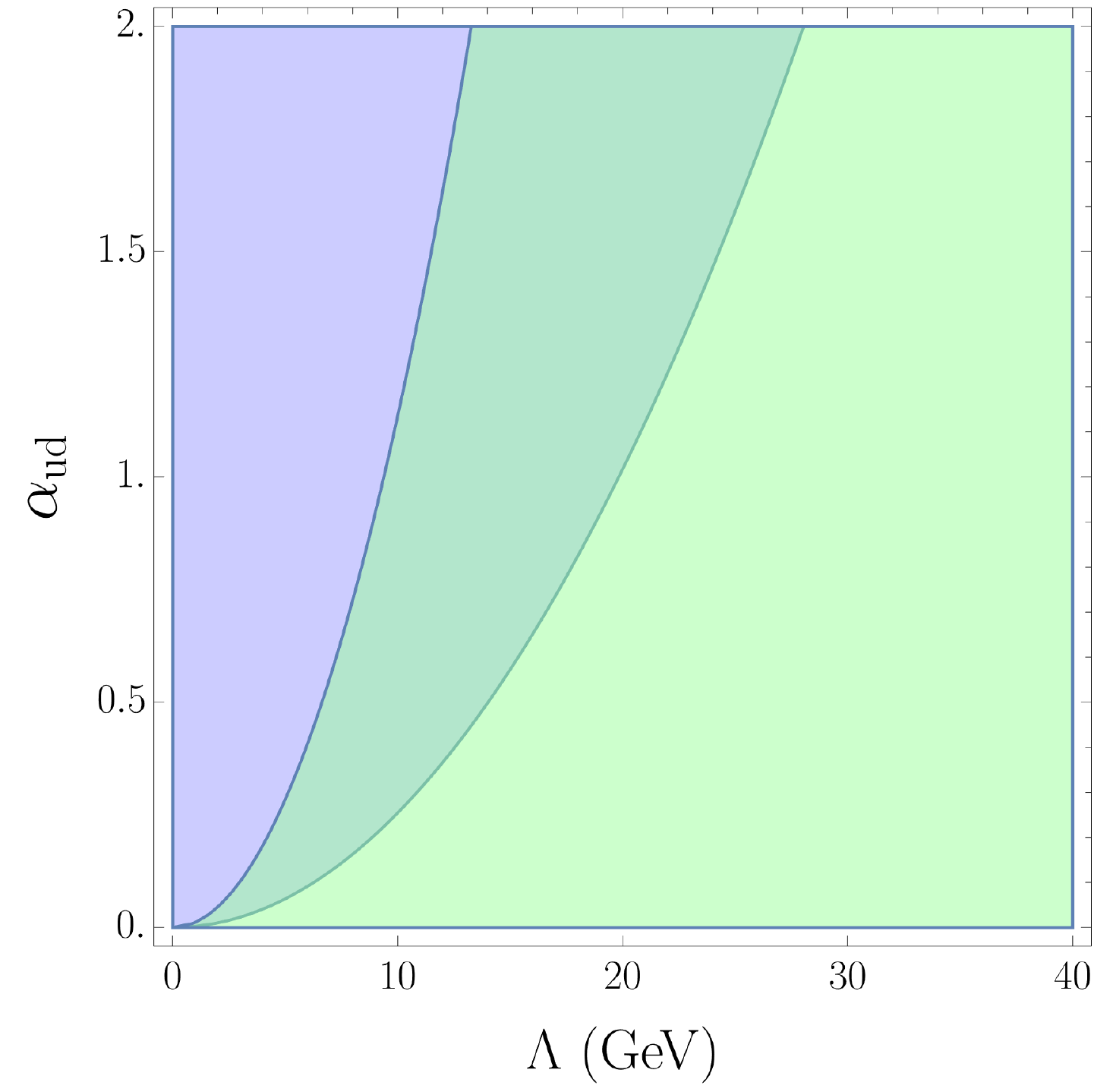}
\includegraphics[scale=.5]{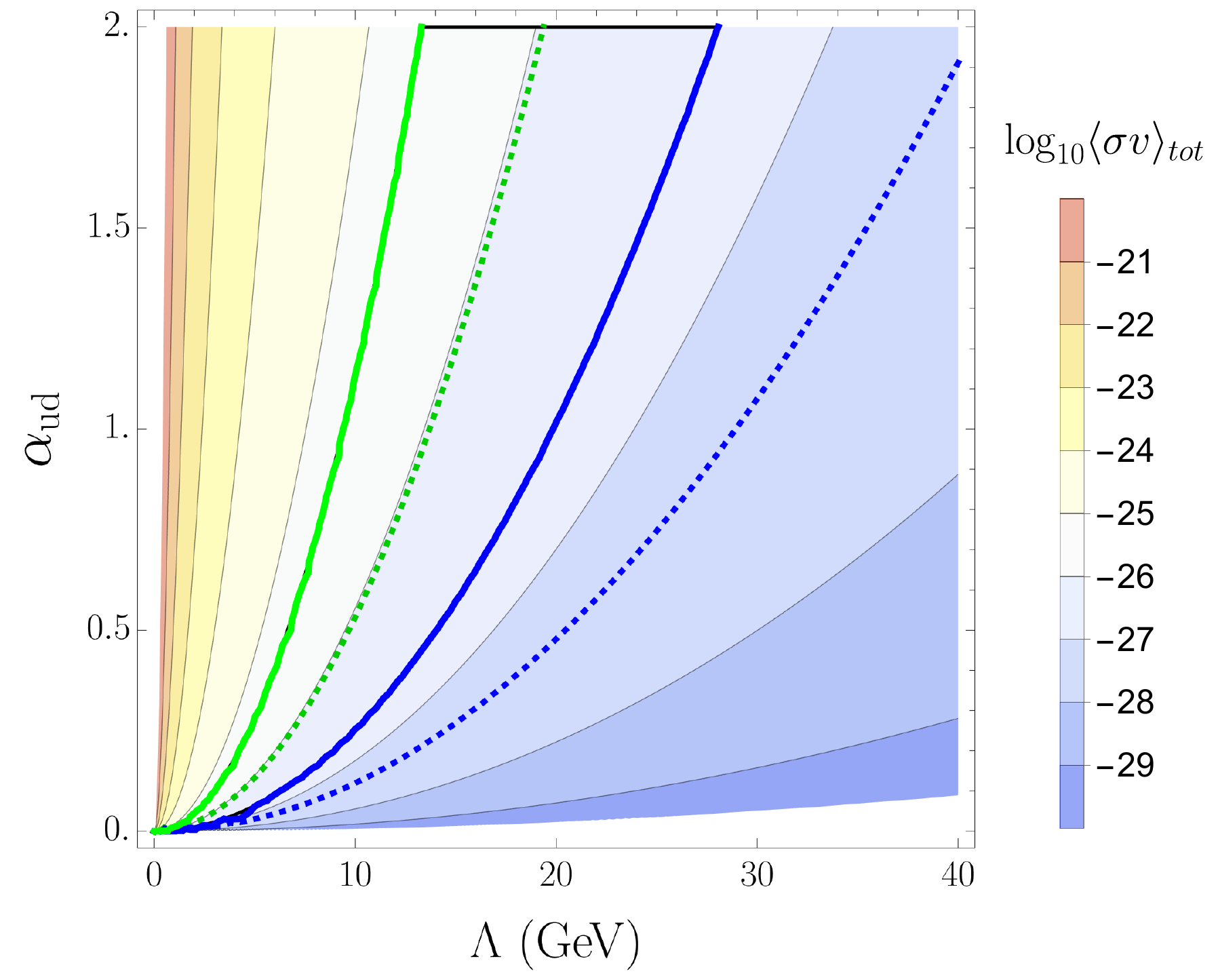}
\caption{{\bf (Left)} Constraints on parameter space induced purely from the results obtained from the chiral Lagrangian. Shown are the regions of parameter space consistent with observations from the diffuse emission (green) and a future $5\sigma$ photon excess from Draco (blue). {\bf (Right)} Density plot of how the total DM annihilation cross section obtained from the chiral Lagrangian varies with the model parameters $\alpha_{ud}, \Lambda$. The two dotted contours are the curves in parameter space which correspond to the total DM annihilation cross sections obtained from the coarse analysis in Section \ref{results}.}
\label{bounds}
\end{figure}

We first consider the constraints imposed on the chiral Lagrangian description of our model. 
Recall that the vertices for each of the kinematically accessible final states are proportional to $\alpha_{ud}/\Lambda^2$, and so the entire photon spectrum will be proportional to $\alpha_{ud}^2/\Lambda^4$ and have no dependence on $\alpha_s$. This is convenient as it allows us to completely explore the parameter space of both $\alpha_{ud}$ and $\Lambda$ to find the combinations which are consistent with current and future observations, which we now do. 

The bounds imposed on the model's parameter space by a future Draco search and current observations of the diffuse emission are shown in Figure \ref{bounds} (left). As mentioned earlier, we have set the center of mass energy to be $\sqrt{s} =2 m_\chi \sim 700~\mev$.
The parameter space consistent with a Draco $5\sigma$ photon excess provides a lower bound on the total DM annihilation cross section, and that the constraints induced by the diffuse emission provides an upper bound. The region overlap is our region of interest, as it is the region of parameter space which must be occupied by the model to qualify as a significant finding. That is, the region of parameter space consistent with both a Draco $5\sigma$ photon excess and current the diffuse emission observations is given by all of the $(\Lambda, \alpha_{ud})$ that fall within the region of overlap between the two regions. This overlap region bound by two contours of constant cross section, defined by

\eq{
  2.10 \times 10^{-27}~ \frac{\cm^3}{\rm s} \ \lesssim \  &\langle \sigma v \rangle_{tot} \ \lesssim  \ 4.18 \times 10^{-26}~ \frac{\cm^3}{\rm s}.
 }


We now consider the constrains imposed on our model that are obtained from our coarse estimates of the cross section \pref{NsMax}. Recall that, as a function of the total DM annihilation cross section, we have found that the predicted photon signal due to DM annihilation can be no larger than 

\eq{
(N_S^{max})_{diff} &= \Theta_{diff} N_c^{tot.} \langle \sigma v\rangle_{tot} =  \langle \sigma v\rangle_{tot} (3.4 \times 10^{33} ~ {\rm cm}^{-3} ~ {\rm {s}}) \label{NsmaxDiff} \\[.4cm]
(N_S^{max})_{draco} &= \Theta_{draco} N_c^{tot.} \langle \sigma v\rangle_{tot} =  \langle \sigma v\rangle_{tot} (5.4 \times 10^{29} ~ {\rm cm}^{-3} ~ {\rm {s}}) \label{NmaxDraco}
}

\noindent Note that these values were obtained with no mention to a chiral Lagrangian, as the only necessary ingredients were a complete list of the possible final states of DM annihilation (which we found from symmetries). Nevertheless, we can still use these quantities to estimate the actual photon signal and thereby estimate the total annihilation cross section as well. We can  then compare the estimates to those obtained from the chiral Lagrangian and provide an error estimate. Such estimates could be useful to future dark matter research for models where the final states are known but a Lagrangian description is either not possible or readily available.

Now, for the diffuse case we again apply the condition that the total predicted photon signal be consistent with the diffuse photon emission. Translating this constraint on the photon signal directly to an upper bound on the total DM annihilation cross section via Equation \ref{NsmaxDiff}, we find that 

\eq{
\langle \sigma v \rangle_{tot} \lesssim 4.17 \times 10^{-27}~ \frac{\cm^3}{\rm s}\label{σminDiff}
}

\noindent is the largest allowed cross section consistent with observations of the diffuse photon emission.

We now consider the lower bound on the total DM annihilation cross section that would be imposed by a $5\sigma$ photon excess from Draco. Now, the minimum photon signal $N_S$ consistent with a 5$\sigma$ photon excess occurs when $N_S = 5\sqrt{N_O^{draco}}$. Now, using Equation \ref{NmaxDraco} to convert this into a bound on the total DM annihilation cross section we find that the lower bound is given by

\eq{
\langle \sigma v \rangle_{tot} \gtrsim 2.10\times 10^{-28}~ \frac{\cm^3}{\rm s}.\label{σminDraco}
}

We visualize these findings in Figure \ref{bounds} (right). Therein we show how the bound obtained via our coarse estimates of the cross sections \pref{σminDiff}, \pref{σminDraco} relates to the cross sections obtained from the more precise description of the chiral Lagrangain. The green contours represent the upper bounds on the cross section by imposing that the dark matter signal be consistent with current observations of the diffuse emission, and the blue contours represent the lower bounds on the annihilation cross section consistent with a Draco 5$\sigma$ photon excess. The solid lines are the predictions obtained from the chiral Lagrangian, and the dotted lines are the predictions obtained from the coarse estimates of the photon signal \pref{NsmaxDiff} and \pref{NmaxDraco}.


Looking at Figure \ref{bounds} (right), we see that our Lagranian-free coarse estimates on the total DM annihilation cross section admit a region in parameter space that is larger and centered around a smaller total cross section than that obtained from the chiral Lagrangian. Specifically, the constraints induced on the total DM annihilation cross section from both analyses are

\eq{
 2.10 \times 10^{-28}~ \frac{\cm^3}{\rm s} \ \lesssim \  &\langle \sigma v \rangle_{tot} \ \lesssim  \ 4.17 \times 10^{-27}~ \frac{\cm^3}{\rm s} \qquad \text{(Coarse Estimates)}\\
  2.10 \times 10^{-27}~ \frac{\cm^3}{\rm s} \ \lesssim \  &\langle \sigma v \rangle_{tot} \ \lesssim  \ 4.18 \times 10^{-26}~ \frac{\cm^3}{\rm s} \qquad \text{(Chiral Lagrangian)}.
}

Note that this particular trend of discrepancies is expected of our estimates. For both cases of Draco and the diffuse spectrum, the estimate on the predicted signal $N_S^{max}$ relates to the predicted signal obtained from the chiral Lagrangian $N_S$ via 

\eq{
N_S  < N_S^{max}
}

\noindent So when we fix $N_S = N_O$ from the chiral Lagrangian, we expect that to correspond to a larger $\langle \sigma v\rangle_{tot}$ than what we obtain when we fix $N_S^{max}=N_O$, since the former is suppressed by the sum over the branching ratios. Therefore, we expect the total annihilation cross section obtained from the coarse estimates to be smaller for both Draco and the diffuse emission, which is exactly what we see in Figure \ref{bounds}.

%

%

\ \\

\section{Conclusions}

In conclusion, we have considered dark matter which communicates to the Standard Model via coupling directly to the quarks as an axial-vector current of the form  $\alpha_q  a_\mu (\barr{q} \gamma^\mu \gamma^5 q)$. We have argued the set of all possible mesonic final states resulting from DM annihilation from kinematics and symmetry considerations. Moreover, we have also found the set of all final states by way of the chiral Lagrangian description of the DM-quark coupling at energies below the confinement scale, which is in agreement with our arguments from symmetry. From these final states we found the photon spectrum for dark matter annihilation.

In this work we have found that current observations of the diffuse photon background emission can already be used to obtain an upper bound on the total dark matter annihilation cross section in a Lagrangian-free manner, as well as the model parameters themselves when using the chiral Lagrangian description. Furthermore, we have found that future observations of dwarf spheroidal galaxies (dSphs) will significantly refine the bounds on the model's parameter space (or possibly discover this model), as it provides a lower bound on the total dark matter annihilation cross section. 

Though this analysis if fairly complete, there are multiple ways one might be able to extend or build from this analysis.  We have considered all allowed final states up to $\Ord(p^4)$ in the chiral Lagrangian. As argued in Section \ref{chiralTheory}, we have therefore included all tree-level contributions to the 3-body final states and have ignored the heavily suppressed $\Ord(p^6)$ correction to the 5-body final states. For completeness, one can choose to expand the $\Ord(p^6)$ chiral Lagrangian and find this contribution. Moreover, one may also choose to include the 1-loop contribution to the 3-body and 5-body final states. However, recall that in Section \ref{spectrum} we showed that a convenient property of the axial-vector case is that the tree-level Feynman rules for the DM contact interactions with the mesons receive no corrections at 1-loop with mesons running the loop. Thus the only possible 1-loop correction to the 3-body or 5-body final states would be with $\barr{\chi}$ and $\chi$ running the loop, which will renormalize the coupling $\alpha_q$. One is invited to calculate this contribution and see how it compares in magnitude to the second order corrections obtained from the chiral Lagrangian. Lastly, one could also increase the mass of DM closer to the confinement scale and see in what way that changes the spectrum and its corresponding bounds on the parameter space and DM annihilation cross section.

\section{Acknowledgments}

We are grateful to Arvind Rajaraman, Jason Kumar and Michael Waterbury for useful and insightful discussions. This work was supported in part by the NSF via grant number DGE-1839285.

%
%
%
%
%
%
%
%
%
%
%
%
%
%
%

\appendix

\section{Boosting the Decay Spectrum}
\label{boostdecayspectrumappendix}
We now describe the general procedure to obtain the boosted spectrum from the
decay spectrum at rest.

We consider
a particle of mass $m$ which
decays into a number of daughter particles, and we assume that the kinematic distribution
of the decay is known in the rest frame of the particle.  Our goal is to determine the
kinematic distribution
in the lab frame, where the parent particle is moving.
We take the parent particle to be traveling  along the $z$-axis, with an energy $E_m$,
corresponding to a Lorentz factor $\gamma=E_m/m$.
We assume that there is no correlation between the direction of the
daughter particle's momentum and the direction of the parent particle's boost.

In the CM frame, the  four-momentum of one of the daughter
particles is $(E', p'\sin\theta', 0, p'\cos\theta')$.
We are given ${dP / dE'}$  in the CM frame; i.e. the probability of obtaining in
the CM frame
a given value of the daughter particle's energy.
In the lab frame,  the four-momentum is $(E, p\sin\theta, 0, p\cos\theta)$.
We are  looking for ${dP / dE}$.

For the daughter particle in the lab frame, we have
\bea
E=\gamma(E'+p' \beta \cos\theta) .
\eea
For any given $E$, this equation has a solution for $\cos \theta$ if $E'$ lies in the range
\bea
\gamma(E-p\beta) \leq E'  \leq \gamma(E+p\beta) .
\eea

The kinematic distribution of the daughter particle in the laboratory frame is then
\bea
\frac{dP(E)}{dE} &=& \frac{1}{2} \int dE'~d\cos \theta \frac{dP(E')}{dE'} \delta
\left(E- \gamma(E' + \beta p' \cos \theta) \right) ,
\nonumber\\
&=& \frac{1}{2} \int_{E_1}^{E_2} dE' \frac{dP(E')}{dE'} {1\over p' \beta\gamma} ,
\eea
where
$E_2=\gamma(E+p\beta)$
and $E_1=\gamma(E-p\beta)$.  This formula allows us to obtain the boosted spectrum from the
decay spectrum at rest.

If we assume that the parent particle itself has a kinematic distribution in the laboratory
frame given by $dN_m / dE_m$, we then find
\bea
\frac{dP(E)}{dE} &=&\frac{1}{2} \int dE_m \frac{dN_m}{dE_m}
\int_{E_1 (E_m)}^{E_2 (E_m)} dE' \frac{dP(E')}{dE'} {1\over p' \beta \gamma} .
\eea
Moreover, if the daughter particle itself decays isotropically to some
tertiary product, one can determine kinematic distribution of this
tertiary product by simply repeating the above process, treating the daughter
particle now as the parent to the tertiary particle.

We can apply this formalism to the case of the $\pi^0$, whose dominant decay is to two photons.
In the rest frame, the photons have back-to-back momenta and
the distribution is $dP /dE'=2\delta(E'-{m_\pi\over 2})$, where
the factor of two accounts for the two photons.
We then find
\bea
\frac{dP}{dE} &=&    \int_{E \gamma (1-\beta)}^{E \gamma (1+\beta)} dE'
\delta \left(E'-\frac{m_\pi}{2} \right){1\over E' \beta\gamma}
\nonumber\\
&=& {2\over \sqrt{E_\pi^2 -m_\pi^2} } \times \left[ \theta\left(E- \frac{m_\pi}{2}
\sqrt{\frac{1-\beta}{1+\beta}} \right)
\theta\left(\frac{m_\pi}{2} \sqrt{\frac{1+\beta}{1-\beta}} -E \right)\right]
\eea
This reproduces the usual box distribution.

If the $\pi^0$ injection spectrum is given by $dN_\pi / dE_\pi$, then we
may express the photon spectrum as~\cite{Boddy:2016hbp}
\bea
  \frac{dN_\gamma}{dE_\gamma} &=&  \int_{\frac{m_\pi}{2} (\frac{2E_\gamma}
	{m_\pi}+\frac{m_\pi}{2E_\gamma})}^\infty dE_\pi
    \left[ \frac{dN_\pi}{dE_\pi} \, \frac{2}{\sqrt{E_\pi^2 - m_\pi^2}}
    \right]~
  \label{eq:GenSpectrum}
\eea
This implies that the photon spectrum is log-symmetric about $m_\pi /2$ with a global maximum at
that point.  Moreover, the spectrum decreases monotonically as the energy either increases or
decreases away from $m_\pi /2$.  We see these features in Figure~\ref{photonSpectrum}.

The last thing which is needed is $dN_\pi / dE_\pi$.  Since the DM only has contact interactions with the pions no boosting is necessary. 

\section{Chiral Lagrangian Technical Details}
\label{chiralAppendix}

At lowest order, the relevant terms in the Lagrangian are the following contact interactions. 

\begin{align}
\Lagr_{\rm contact}^{(2)}  &= \frac{\sqrt{2}}{3}  \alpha_{du} \beta  a_{\mu } \bigg[ 
\pi^0 \pi ^- \del _{\mu }\pi ^+ + \pi ^+ \pi^0 \del _{\mu }\pi ^- -2 \pi ^+\pi ^- \del _{\mu }\pi^0 
\bigg]\\
& -\frac{\alpha_{du} \beta ^3 }{15 \sqrt{2}}a_{\mu }  ( \pi^0 )^2 \bigg[
\pi^0 \pi ^- \del _{\mu }\pi ^+ + \pi ^+ \pi^0 \del _{\mu }\pi ^- -2 \pi ^+ \pi ^- \del _{\mu }\pi^0
\bigg]\\
& 
-\frac{ \sqrt{2}}{15} \alpha_{du} \beta ^3  a_{\mu } \pi ^- \pi ^+  \bigg[ 
\pi^0 \pi ^- \del _{\mu }\pi ^+ + \pi ^+  \pi^0 \del _{\mu }\pi ^- -2 \pi ^+ \pi ^- \del _{\mu }\pi^0
\bigg]
\end{align} 

\noindent Now, we decompose the $\Ord(p^4)$ Lagrangian $\Lagr^{(4)}_{\rm contact}$ as 

\begin{align}
\Lagr^{(4)}_{\rm contact} & = \Lagr^{(4)}_{0+-} + \Lagr^{(4)}_{000}+ \Lagr^{(4)}_{000+-} + \Lagr^{(4)}_{0++--}
\end{align}

\noindent Expanding the chiral Lagrangian into its mesonic fields we obtain corrections to the $\pi^0 \pi^+ \pi^-,\pi^0\pi^+\pi^- \pi^+ \pi^-, \pi^0\pi^0\pi^0 \pi^+ \pi^-$ processes and the lowest order contribution to the $\pi^0\pi^0 \pi^0$ vertex. 

\begin{align}
\Lagr_{\rm 0+-}^{(4)}& = 
\frac{\alpha_{du} \beta ^3}{\sqrt{2}} L_9   \del _{\nu }a_{\mu }\bigg[
\pi ^-  \del _{\mu }\pi^0 \del _{\nu }\pi ^++\pi ^+  \del _{\mu }\pi^0 \del _{\nu }\pi ^- - \pi ^+ \del _{\mu }\pi ^- \del _{\nu }\pi^0- \pi ^- \del _{\mu }\pi ^+ \del _{\nu }\pi^0
 \bigg] + \mu \leftrightarrow \nu
\end{align}

\begin{align}
\Lagr_{000} &= \sqrt{2} \beta ^3 \alpha_{du} \left(2 L_1+2 L_2+L_3\right) \del _{\mu }\pi^0 \del _{\nu }\pi^0  \Big(a_{\mu } \del _{\nu }\pi^0+a_{\nu } \del _{\mu }\pi^0\Big) 
\end{align}

\begin{align}
\Lagr^{(4)}_{000+-}& = \frac{\alpha_{du} \beta ^5}{6 \sqrt{2}} a_{\mu } \bigg[
 4 \left(2 L_1+L_2+L_3\right)\pi _0 \pi ^-  \del _{\mu }\pi ^+ \left(\del _{\nu }\pi _0\right){}^2+4  \left(2 L_1+L_2+L_3\right) \pi _0 \pi ^+ \del _{\mu }\pi ^- \left(\del _{\nu }\pi _0\right){}^2\\
 &-12  \left(2 L_1+2 L_2+L_3\right) \pi ^- \pi ^+\del _{\mu }\pi _0 \left(\del _{\nu }\pi _0\right){}^2-4  L_2 (\pi^0)^2 \del _{\mu }\pi ^+ \del _{\nu }\pi ^- \del _{\nu }\pi _0-4  L_2  (\pi^0)^2 \del _{\mu }\pi ^- \del _{\nu }\pi ^+ \del _{\nu }\pi _0\\
 &+4 \left(2 L_1+3 L_2+L_3\right) \pi _0 \pi ^- \del _{\mu }\pi _0 \del _{\nu }\pi ^+ \del _{\nu }\pi _0+4 \left(2 L_1+3 L_2+L_3\right)  \pi _0 \pi ^+ \del _{\mu }\pi _0 \del _{\nu }\pi ^- \del _{\nu }\pi _0\\
 &-4  \left(2 L_1+L_3\right) (\pi^0)^2 \del _{\mu }\pi _0 \del _{\nu }\pi ^- \del _{\nu }\pi ^+
\bigg] + \mu \leftrightarrow \nu 
\end{align}

\begin{align}
\Lagr^{(4)}_{0++--} &= 
\frac{\alpha_{du}\beta ^5}{3 \sqrt{2}} a_{\mu } \bigg[ 
 \left(2 L_1+L_3\right) \left(\pi ^+\right)^2 \del _{\mu }\pi^0\left(\del _{\nu }\pi ^-\right){}^2+4 L_2 \pi^0\pi ^+  \del _{\mu }\pi ^+ \left(\del _{\nu }\pi ^-\right){}^2 + 2 L_2 \left(\pi ^+\right)^2  \del _{\mu }\pi ^- \del _{\nu }\pi^0\del _{\nu }\pi ^- \\
&+4  \left(2 L_1+L_2+L_3\right) \pi^0\pi ^- \del _{\mu }\pi ^+ \del _{\nu }\pi ^+ \del _{\nu }\pi ^- +4  \left(2 L_1+L_2+L_3\right) \pi^0\pi ^+ \del _{\mu }\pi ^- \del _{\nu }\pi ^+ \del _{\nu }\pi ^- \\
&-10 L_2 \pi ^- \pi ^+  \del _{\mu }\pi ^+ \del _{\nu }\pi^0\del _{\nu }\pi ^- -10 \left(2 L_1+L_3\right)  \pi ^- \pi ^+ \del _{\mu }\pi^0\del _{\nu }\pi ^+ \del _{\nu }\pi ^- +2 L_2 \left(\pi ^-\right)^2  \del _{\mu }\pi ^+ \del _{\nu }\pi^0\del _{\nu }\pi ^+ \\
&+ \left(2 L_1+L_3\right) \left(\pi ^-\right)^2  \del _{\mu }\pi^0\left(\del _{\nu }\pi ^+\right){}^2 +4 L_2 \pi^0\pi ^-  \del _{\mu }\pi ^- \left(\del _{\nu }\pi ^+\right){}^2-10 L_2\pi ^- \pi ^+  \del _{\mu }\pi ^- \del _{\nu }\pi^0\del _{\nu }\pi ^+
\bigg] + \mu \leftrightarrow \nu.
\end{align}

Throughout this paper, we make use of the definitions 

\eq{
\alpha_{ij} = \alpha_{i} - \alpha_{j} &\qquad \text{for}  \qquad   i,j \in \{ u, d, s \}\\[.3cm]
\beta &= \frac{\sqrt{2}}{f}.
}

\end{document}